\documentclass[10pt]{sig-alternate-ipsn13}

\usepackage{graphicx,changepage}
\usepackage{float}
\usepackage{color}
\usepackage{xcolor}
\usepackage{xspace}
\usepackage{amsmath}
\usepackage{amssymb}
\usepackage{multirow}
\usepackage{multicol}
\usepackage{comment}
\usepackage{framed}
\usepackage{algorithm}
\usepackage{program}
\usepackage{listings}
\usepackage[font=bf,skip= 3pt,size = small]{caption}
\usepackage{times}    
\usepackage{subfigure}

\floatstyle{ruled}
\newfloat{grammar}{thp}{lop}
\floatname{grammar}{Rule Syntax}

\floatstyle{ruled}
\newfloat{property}{thp}{lop}
\floatname{property}{Property}

\numberwithin{algorithm}{section}
\newfloat{program}{thp}{lop}
\floatname{program}{Pseudocode}

\definecolor{shadecolor}{gray}{0.9}
\definecolor{light-gray}{gray}{0.9}

\newcommand{\tab}{\hspace*{2em}}

\def\pprw{8.5in}
\def\pprh{11in}

\makeatletter
 \let\@copyrightspace\relax
\makeatother

\begin{document}

\title{Model-Driven Design of Clinical Guidance Systems}

\numberofauthors{3}
\author{
\alignauthor
Maryam Rahmaniheris\\
       \affaddr{Dep. of Computer Science}\\
       \affaddr{University of Illinois}\\
       \affaddr{Urbana, IL 61801, USA}\\
       \email{rahmani1@illinois.edu}
\alignauthor
Yu Jiang\\
       \affaddr{Dep. of Computer Science}\\
       \affaddr{University of Illinois}\\
       \affaddr{Urbana, IL 61801, USA}\\
       \email{jy1989@illinois.edu}
\alignauthor
Lui Sha\\
       \affaddr{Dep. of Computer Science}\\
       \affaddr{University of Illinois}\\
       \affaddr{Urbana, IL 61801, USA}\\
       \email{lrs@illinois.edu}
}


\maketitle


\begin{abstract}

Clinical guidance systems have been widely adopted to help medical staffs to avoid preventable medical errors such as delay in diagnosis, treatment or untended deviations from best practice guidelines. However, because patient condition changes rapidly
and medical staffs are usually overloaded in acute care setting,
how to ensure the correctness of the system reaction to those rapid changes and managing interaction with physicians remains challenging. In this paper, we propose a domain-specific model-driven design approach to address these challenges for designing clinical guidance systems. Firstly, we translate the relevant medical knowledge described in generic literature and different best practice guidelines into a set of executable and indirectly verifiable finite state machine models. We introduce an organ-centric paradigm to construct clinical models, and also develop a physician model to track physician-system interactions and deviations. Secondly, for verification of the compositional system model, we translate the model into timed automata, based on which, we formalize a set of clinical and system safety requirements as computation tree logic(CTL) formulas and use the UPPAAL model checking tool to formally verify those requirement. In this way, the correctness of the model can be mathematically proved. Finally, we can automatically generate the executable code from the verified model using the corresponding code generation tools for finite state machine. For evaluation, we apply the approach to the design of clinical guidance systems for cardiac arrest. The generated code can be deployed and interact with existing guidance systems.

\end{abstract}

\section{Introduction} \label{sec:introduction}

According to the Institute of Medicine, close to 100,000 safety-related medical incidents happens each year in US due to \textit{preventable} medical errors \cite{iom1}.
These errors are often caused by delayed diagnosis, delayed or ineffective therapeutic interventions, and unintended deviation from the best practice guidelines.
These situations may occur more often in acute care settings, 
where the medical staffs are overloaded, and must make quick decisions based on the best available evidence.

An integrated clinical guidance system can reduce such medical errors in acute care by helping medical staff track and assess patient state more accurately and adapt her care plan according to the best practice guidelines. Clinical guidance systems can greatly affect patient safety,and therefore, it is critical to validate and formally verify their correctness. However, medical information is considerably vast and complex. Moreover, the functional complexity of modern clinical systems has increased over the years. This makes it difficult to ensure the correctness of the encoded medical knowledge in the system.

Traditionally, such systems may be developed by directly implementing the clinical monitoring and treatment protocols in procedural style using a high level language such as Java. The proof of correctness process can become complicated. Moreover, it would be greatly valuable to have physician in the loop to validate the clinical correctness of the encoded medical knowledge. If implemented using a high level language such as Java, patient state representation and other encoded clinical knowledge could not be easily validated by physicians. Therefore, it is critical to have representations at the same (or as close as possible to) level of abstraction, which medical knowledge and reasoning are known to physicians.

\begin{figure} [!htbp]
	\center
	\includegraphics[width=0.99\columnwidth]{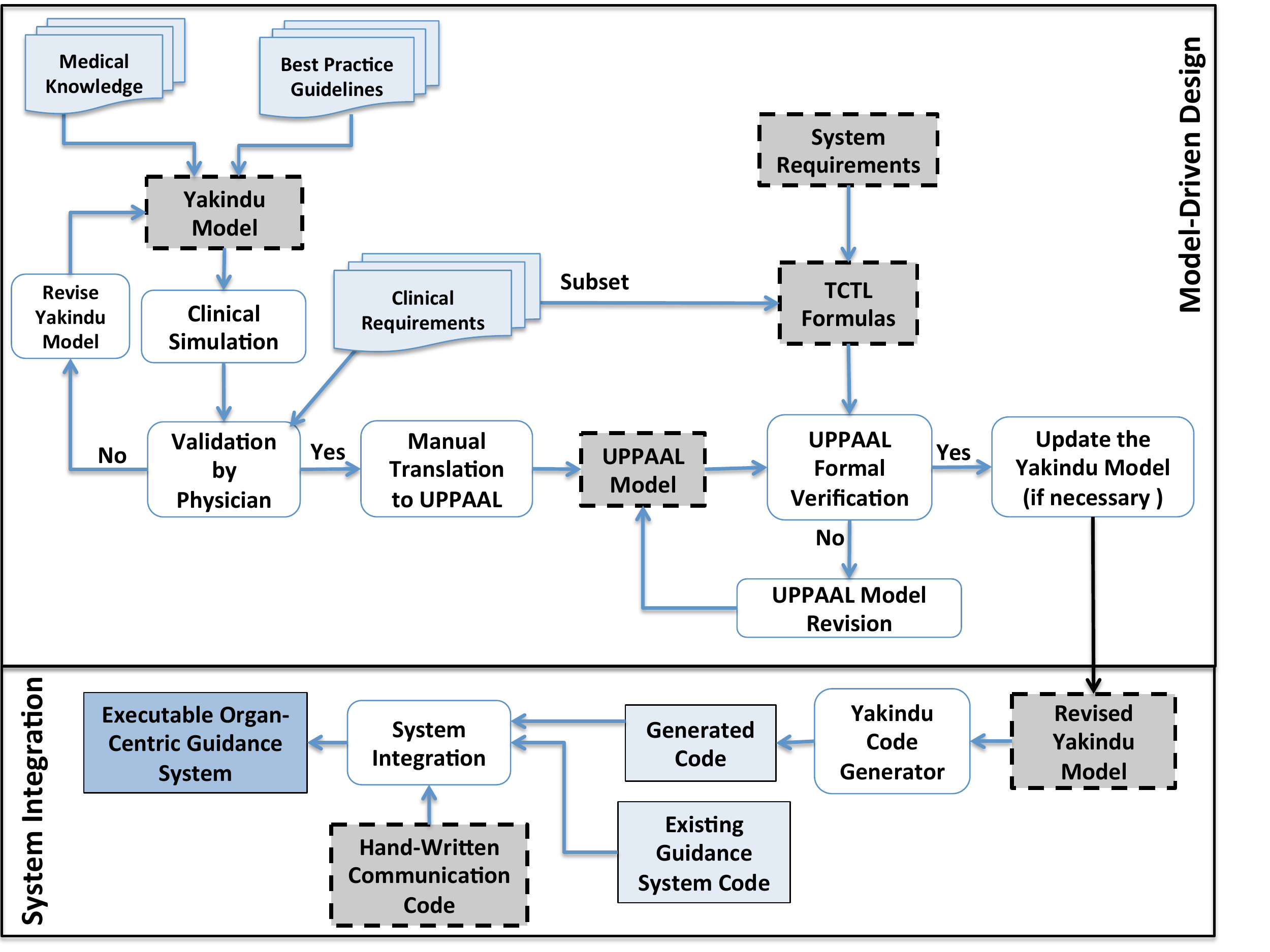}
	\caption{Model-Driven Design for Clinical Guidance System }
	\label{fig:approach}
\end{figure}

\textbf{\textit{Proposed Approach}:} In this paper, we propose a domain-specific model-driven approach for designing clinical guidance system, in which we develop a cyber-medical-human. Figure \ref{fig:approach} shows the overview of our approach, which includes the following main steps: \textbf{1)} We first translate the generic medical knowledge into a set of communicating finite state machines. We propose an organ-centric paradigm, in which we model patient state as a network of disease and organ state machines. We use the finite state machine formalism to develop a best practice manager model that encodes all the monitoring and treatment guidelines. A compositional automata model is also proposed to handle and track the physician interaction with the system. We use an open source tool, called Yakindu, to develop the models. The tool provides an integrated modeling environment and development based on the concept of statecharts. The tool also allows simulation of the developed model to validate its dynamic behavior \cite{yakindu}. \textbf{2)} A set of clinical requirements have been drafted in collaboration with our physician collaborators. Physician can directly look at these organ models and cross reference with the specified clinical requirements. Note that, the developed models are at same level of abstraction the medical knowledge and reasoning are known to physician. At our proposed level of abstraction, physicians with minimal IT background can easily understand the models and review the design. Furthermore, Yakindu simulation of executable organ models allows dynamic clinical validation. We then apply the necessary changes, as suggested by physician(s), and re-validate the refined model. \textbf{3)} The finalized Yakindu model is manually translated to UPPAAL for exhaustive verification of clinical and system requirements. The clinical requirements drafted previously are formalized as computation tree logic formulas and along side the system requirements are verified using UPPAAL. Note that any necessary modification to UPPAAL model are reflected back to the Yakindu Model. 4) As the last step, the Java code is generated automatically from the validated and verified Yakindu model and is integrated with the code from an existing guidance system, using some handwritten communication code.

\textbf{\textit{Main Contribution}:} Overall contributions are:

\begin{enumerate}
	\item Proposing a model-driven approach for the design of clinical guidance systems, where a cyber-medical-human model has been developed that facilitates clinical validation by physician. The proposed model also includes a compositional physician model, which tracks and manages the physician interaction with the system. We have formulated a set of clinical and system requirements to verify against the developed models.
	
	\item Using real case study to show how the models can be constructed and then integrated into an existing guidance system.
	
\end{enumerate}

\textit{\textbf{Paper Organization:}} An overview of the clinical guidance system is presented in Section \ref{sec:overview}. We present the detail of our model-driven approach in Section \ref{sec:model}. The details of our proposed cyber-medical-model is discussed in Section \ref{construction}; clinical validation and verification of the models are addressed in Sections \ref{validation} and \ref{verification}. We demonstrate how the developed models are integrated in the clinical guidance system in Section \ref{sec:guidance}. We further elaborate our approach using the cardiac arrest case study in Section \ref{sec:case-studies}. Finally, the related work and our concluding remarks are presented in Sections \ref{sec:related} and \ref{sec:conclusion}, respectively.

\section{System Overview} \label{sec:overview}

\begin{figure} [!htbp]
	\center
	\includegraphics[width=0.99\columnwidth]{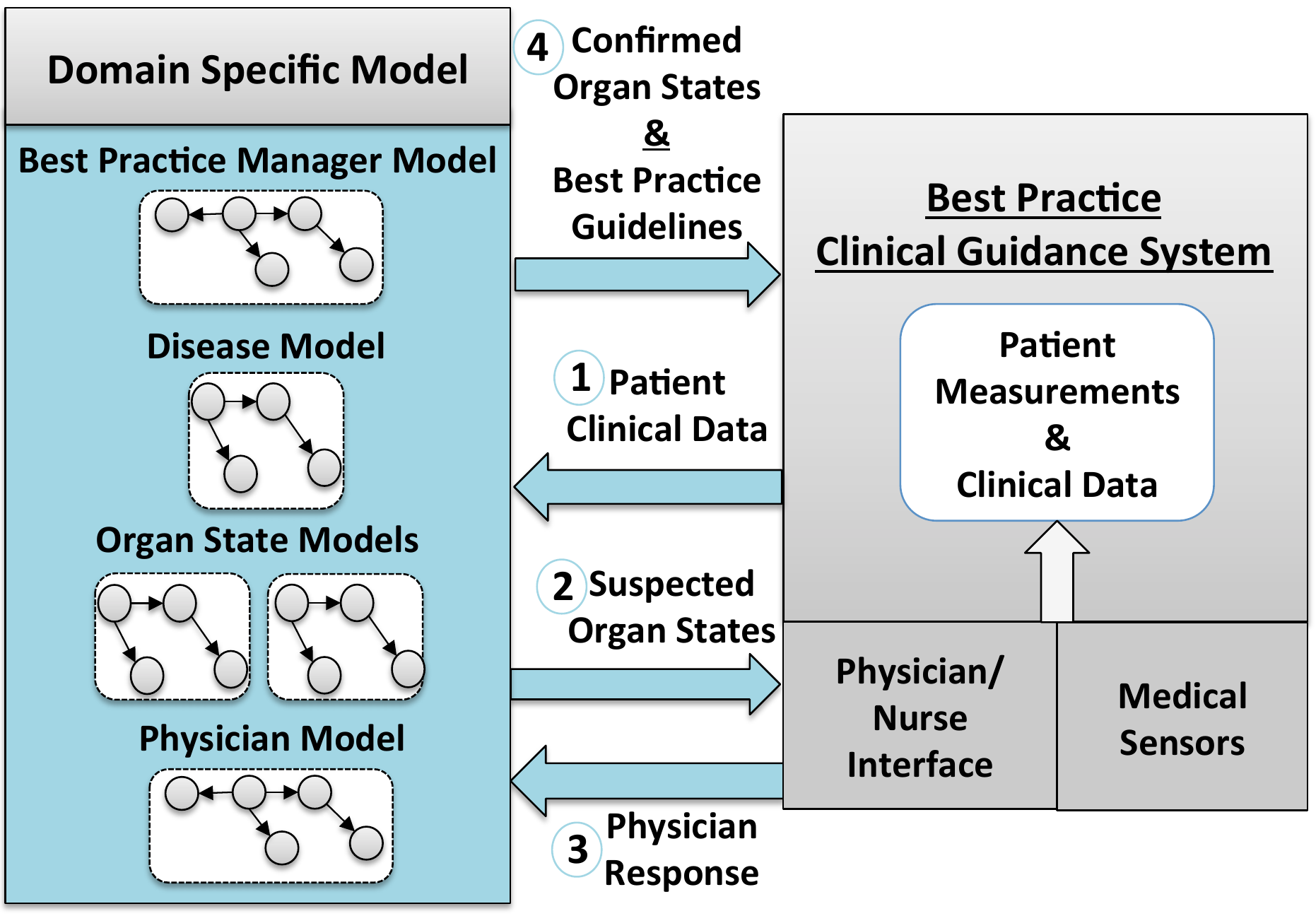}
	\caption{System Overview}
	\label{fig:overview}
\end{figure}

Figure \ref{fig:overview} shows the overall architecture of the proposed approach. Our domain-specific model is at the heart of this architecture. The disease and organ models, developed using communicating  state machine formalism, allow us to keep track of multiple organ states changing concurrently. As new patient data is received, each organ state machine is updated to reflect the most recent patient condition. The model also includes a physician manager, which is responsible for interaction with the attending physician and keeping track of his/her response. The physician can confirm or put a hold on the patient state suggested by our system. The best practice model encodes the monitoring and treatment guidelines. We only consider the diseases and acute care scenarios, where best practice algorithm and guidelines exist. One example is the ACLS (Advanced Cardiac Life Support) algorithm provided by American Heart Association (\textit{AHA}) to reverse cardiac arrest.

\section{Model-Driven Approach} \label{sec:model}

As shown in Figure \ref{fig:approach}, developing efficient cyber-medical models is the first step in our model-driven approach for design of clinical guidance systems. We first translate the relevant medical knowledge described in generic literature and different best practice guidelines into a set of executable models, using finite state machine formalism. We have chosen to use an organ-based pathophysiological paradigm for developing the models. We have also developed a best practice manager that encodes the treatment and monitoring guidelines. On the other hand, the role of human in medical system can be more critical compared to most of other \textit{cyber-physical human systems (CPHS)}. Moreover, the complexity and uncertainty of medicine sometimes makes definitive decisions difficult. Therefore, it is critical to account for the human (specifically, physician) as an important aspect of the system. In this paper we propose a compositional model to record and track physician response to the information provided by our system. In the following subsections, we discuss the details of this paradigm and how the models are constructed. However, to explain the model construction more clearly we first look at an illustrative example. More details on this case study is presented in Section \ref{sec:case-studies}. 

\subsection{Illustrative Example}

We use the case of cardiac arrest scenario to illustrate the proposed model construction procedure. Cardiac arrest is initially the devastating cessation of cardiac activity, which is caused by the loss of heart's electrical and muscular pumping function. The cardiovascular organ system is the main organ affected by this disease. In cardiac arrest, the heart rhythm may change and include abnormal or irregular heart rhythms or waveforms (arrhythmias). Furthermore, The insufficient blood circulation in the lungs impedes the $CO_{2}$ and $O_{2}$ exchange resulting in $CO_{2}$ accumulation in the blood, which may be related to a rise in blood acidity level indicated by decrease in blood $pH$ level. The abnormal $pH$ is revealed in arterial blood gas (ABG) test. These are two examples of pathophysiological changes in cardiovascular organ system caused by cardiac arrest. In addition to the cardiovascular organ system, as resuscitation efforts continue, it may be found that the lung (pulmonary) and kidney (renal) organ systems are also affected

Cardiac arrest can lead to death within minutes. The guideline provided by American Heart Association (\textit{AHA}) to reverse the cardiac arrest includes cardiopulmonary resuscitation (\textit{CPR}), using the defibrillator device to shock the heart by delivering a therapeutic dose of electrical energy to the heart, and administration of intravenous drugs such as epinephrine \cite{acls}. Epinephrine is a vessel constrictor that stimulates the heart and increases the arterial blood pressure.

\begin{figure*}[!t] 
	\centering
	\subfigure[Best Practice Automaton]{\label{fig:arrhythmiaBP} 	\includegraphics[scale=0.4]{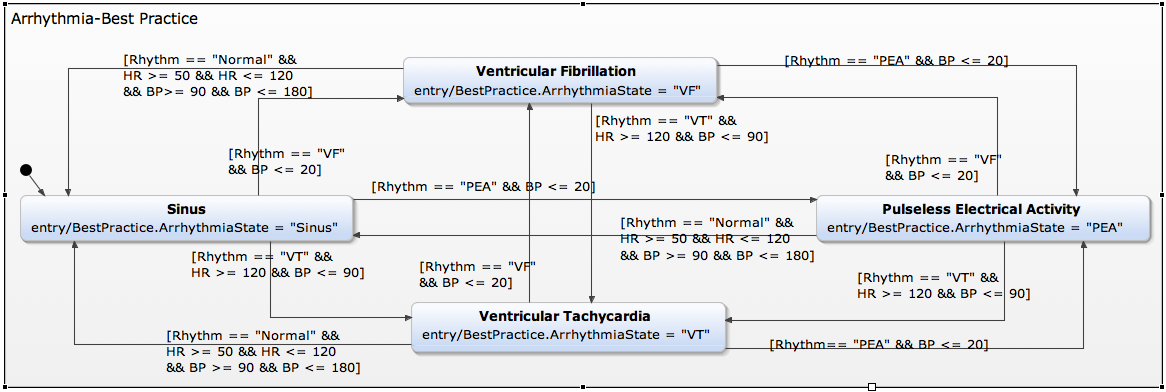}}
	\subfigure[Physician Automaton] {\label{fig:arrhythmiaPhysician} 	\includegraphics[scale=0.4]{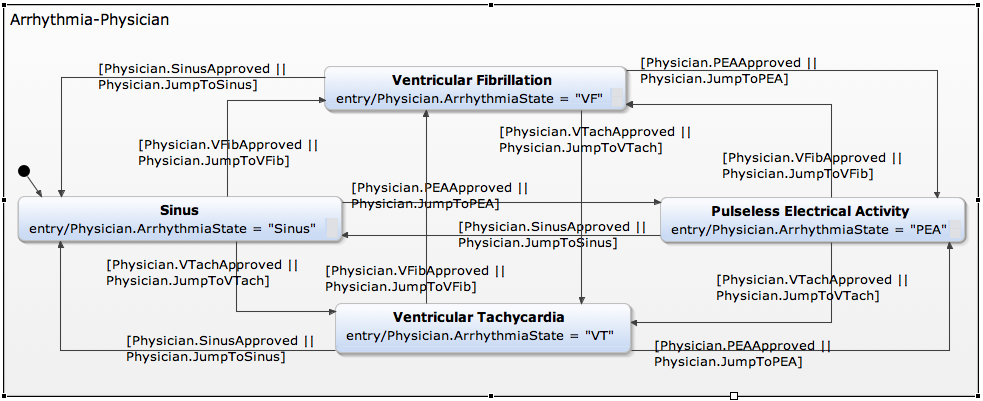}} 
	\caption{Manually Constructed Arrhythmia Automata in Yakindu}
	\label{fig:arrhythmia}
\end{figure*}

\subsection{Model Construction} \label{construction}

\subsubsection{Organ-Centric Pathophysiological Model}

We have chosen to translate medical knowledge to executable models using an organ-based pathophysiological\\ paradigm. We model the changes in patient state according to the pathophysiology of affected organ systems. Different diseases may impair one or more of the physiological functions performed by different organ systems. Pathophysiology describes the process of such change in physiological measurements and organ states throughout disease progress. Multiple pathophysiological processes must often be tracked for each disease. The main reasons behind developing our models according to this paradigm are as follows:

\begin{itemize}
	
	\item Physicians are taught body organ systems and structural relationships within the body. Moreover, the pathophysiological description of disease processes is how medical knowledge is commonly known by the physicians \cite{berlin2, berlin3}. Therefore, the developed models would be at the same abstraction level that medical knowledge and reasoning are known to physicians. Consequently, organizing patient information according to organ-based pathophysiological models can make clinical validation easier and more effective. Physicians may directly look at the organ models and cross reference with the requirements.


	\item Physicians often describe  \textit{monitoring and therapeutic requirements for different organ states}. For example, ACLS algorithm indicates that \textit{shock} must be considered when shockable rhythm is confirmed. This is a treatment guideline for arrhythmia, which is an insufficiency of cardiovascular organ system. Therefore, the proposed modeling paradigm could also facilitate the design of any clinical guidance system that encodes the best practice monitoring and treatment guidelines.

\end{itemize}

We use finite state machine formalism to model pathophysiological processes of different organ systems. Each automaton is a labeled transition system. The nodes represent different stages/types organ insufficiency. The state variables may include physiological measurements and lab values. The definition of each state and the condition enabling transitions between the states are extracted from medical literature and clinical best practice guidelines.

In this section we discuss our modeling methodology and the steps we have taken during the modeling process to make the clinical validation less complex and more straightforward. In the process of developing our pathophysiological organ models, we have gone through multiple iterations of design, validation and close discussion with our ICU physician collaborators. At each iteration a set of modeling guidelines were generated, which were used to update the pathophysiological models for the next iteration. The following modeling guidelines are generated as a result of this whole process:

\textbf{G1. Considering all relevant pathophysiological models in the best practice guideline:} As mentioned in the previous section, each disease can cause several abnormalities and insufficiencies in different organ systems. For example, in cardiac arrest, the main affected organ is the cardiovascular system. The effect mainly includes abnormal heart rhythm and blood acidity and gas levels. However, other organs might also be affected by cardiac arrest. For example, while medical staff focus on treating arrhythmia and injecting various drugs, such as epinephrine, patient may develop renal insufficiency. If the medical staff do not notice the new condition and continue giving the same drugs with the same dosage, patient's kidney may become severely insufficient or even fail. Therefore, we must include all the organ and insufficiency category models relevant to the disease under study. For cardiac arrest, in addition to arrhythmia and blood gas imbalance state machines, our model also includes a renal insufficiency state machine to demonstrate the multi-organ interactions.

\textbf{G2. Determining the correct level of concurrency:} Another important modeling choice is modeling a pathophysiological process by a single state machine vs. multiple concurrent state machines. For each disease, several organ insufficiencies may happen simultaneously. To keep track of all abnormalities, we model each insufficiency as a  separate state machine. This also increases the modularity and prevent any critical patient state change from being overlooked. We also observe that, there are measurements, which physicians often look at as a group and interpret together to have a better assessment of patient state. We decided to model all such measurements in one single process represented by one single state machine. An example of this case, is the urine output, creatinine and potassium Physicians often evaluate these values together to conclude the degree of renal insufficiency. Putting the two guidelines together, for the cardiac arrest case study we have designed a renal insufficiency machine to keep track of kidney state, which can change in parallel with cardiovascular organ functions. We use these three measurements to define the different states of renal insufficiency in the kidney state machine.

\textbf{G3. Choosing the correct level of abstraction for each model:} The abstraction level for each automaton is derived from the best practice algorithm for the disease in cooperation with the physicians. For example, in the case of cardiac arrest, most guidelines focus on cardiovascular system as the main affected organ. Therefore, more detailed information regarding cardiovascular abnormalities such as arrhythmia, blood gas imbalance and etc. may be codified in the model. The organ insufficiency states are defined as a function of physiological measurements. This information is retrieved from medical literature and/or given by our physician collaborators. 

\textbf{G4. Correlating relevant clinical data:} Most current medical systems raise alerts based on single measurements, which creates considerable false positive alerts. Identifying the relation between measurements and structuring them according to organ pathophysiology add a layer of intelligence to patient state representation. For example, we consider heart rate and blood pressure in addition to the EKG rhythm to help medical staff keep track of the state of arrhythmia in a cardiac arrest patient. This enables us to detect inconsistencies between correlated measurements and help medical staff with a more precise assessment of patient state. Following the proceeding example, if the EKG rhythm indicates $PEA$ but the systolic blood pressure reading is above $20 mmHg$, our model recognizes this inconsistency and informs the medical staff of the issue. In this example, there might be a problem with blood pressure cuff or one of the EKG leads might be loose. Bringing such inconsistencies to the attention of physician can reduce the likelihood of administering the inappropriate treatment due to imprecise patient state evaluation. 

\textbf{G5. Defining clinically sound and useful states:} In the previous two guidelines, we discuss the number of parallel state machines, and the measurements to be included in each machine. The next step is to define each state as a function of the chosen physiological measurements. The definition of each state must be clinically sound; however, defining a large number of states for each machine can make the model very complex and unmanageable. Moreover, not all the changes in physiological measurements are relevant or useful for a given disease. Providing too much information and too many alerts to the medical staff may create confusion and increase their mental overload, which is what our system aims to prevent in the first place. Therefore, we have used the notion of \textit{actionable states}. We define a new state when there is a need for new monitoring and therapeutic plan. The actionable states can be extracted from the requirements. Note that each requirement specify the organ states, which trigger an action from the medical staff.

\textbf{G6. Defining state transitions}: Similar to the definition of states in each machine, defining a large number of transitions can make the model complex and less understandable. Moreover, the process of model development becomes more error prone. The key in identifying the correct and optimal number of transitions is considering the medical knowledge underlying our models. Development and similarly resolution of organ failure may take some time. Therefore, when continuous physiological variables are used to define different insufficiency states, we may not need to create an edge between every insufficiency states. The exception may be the initial state; since we do not know the patient state when we start, we can define a transition between the initial state and every other insufficiency states. This can remove the any potential latency in patient state tracking associated with the model stabilization. If we do not have the transitions edges from the initial state, when the system starts the models might need a few cycles to stabilize and reflect the current patient state.


Following the modeling guidelines, \textit{G1-G3}, and after a series of discussions with our ICU physician collaboratives, we have come up with a set of organ insufficiency models for cardiac arrest:  the \textit{cardiac arrhythmia}, \textit{blood gas imbalance (BGI)}, and \textit{renal insufficiency} automata. Let us take a closer look at the arrhythmia machine as an example. As you can see in Figure \ref{fig:arrhythmiaBP}, the model represent different types of heart rhythm irregularity. Let us assume the current readings indicate no meaningful heart rate reading and an EKG rhythm that is not a flat line but does not produce a pulse. Therefore, the \textit{arrhythmia} machine transitions to \textit{pulseless electrical activity} state. As a result, the monitoring system using these developed organ models can detect new organ state and notify the medical staff of the changes in patient state. Different states of arrhythmia may require different treatments, and therefore are considered actionable states (\textit{G5}). Furthermore, since the EKG rhyme is not considered a continuous measurement, following \textit{G6}, we created transition edges between all the states.  

As mentioned before, the state variables and physiological conditions enabling state transitions for each automaton come from the best practice guidelines. In the arrhythmia model, the EKG rhythm is the main state variable; however, following \textit{G4} and through discussion with our physician collaborators, blood pressure and heart rate measurements are used in addition to EKG rhythm for a better assessment of the arrhythmia state. Note that the cut-off thresholds for the blood pressure and heart rate, shown in Figure \ref{fig:arrhythmiaBP}, are for average patients. We use numbers for better readability. \textit{In fact, such threshold values are defines as parameters to our model, which can be configured by the attending physician before the system starts}. In the next section we present our physician-in-the-loop paradigm form run-time interaction with the attending physician.

\subsubsection{Human Model} \label{sec:physician} 

At times patient states presented by our system may differ from physician understanding and they may deviate from the best practice guidelines suggested by our system. A source of this inconsistency, should it occur, is that there may be information used by the physician to make a determination but not available to our system, or the inconsistency cab be the result of unintended deviation. The physician my override the best practice guidelines at anytime. However, any deviations from the best practice guidelines must be tracked and recorded for offline analysis.   

We propose a compositional automata model to handle and track the physician interaction with the system. For each patient state suggested by our model, the response from the physician regarding the new patient state is received and processed by a physician automata. We dedicate a separate physician automata for each organ automata included in the disease model. Figure \ref{fig:arrhythmiaPhysician}, for example, shows the physician automata for \textit{cardiac arrhythmia}, which represents the physician's belief on the current state of the corresponding pathophysiological process. If the physician does not approve the organ state suggested by the system, the physician automata does not transition to the suggested organ state. Moreover, at each time physician may decide to transition (\textit{jump}) to a new state independent of the organ states suggested by our system. As seen in Figure \ref{fig:arrhythmia}, the physician and best practice organ automata for each pathophysiological process have identical structures but can have different run-time states. This compositional approach results in more modular approach, in which the organ and physician models can be validated and verified independently.

\begin{program}[!ht]
	
	{\small
		\begin{lstlisting}[breaklines, mathescape, numbers=left, xleftmargin=15pt]
		input 1: $Organ Automata$ $\textbf{\textit{OA}}=(OA_{1},OA_{2},\dots,OA_{n})$ 
		input 2: $Physician Automata$ $\textbf{\textit{PA}}=(PA_{1},PA_{2},\dots,PA_{n})$  
		input 3: $Patient State$ $\textbf{\textit{S}}= (S_{1},S_{2},\dots,S_{n})$
		input 4: $Physician Belief$ $\textbf{\textit{B}} = (B_{1},B_{2},\dots,B_{n})$
		input 5: $Best Practice Guidelines$ $\textbf{\textit{G}}$
		input 6,7: Short Timer $\textbf{\textit{T}}_{\textit{\textbf{short}}}$, Long Timer $\textbf{\textit{T}}_{\textbf{\textit{long}}}$ 
		
		$UPDATE();$
		$\textbf{if}$ $S \neq B$
		$\tab$ $Send$ $\textbf{\textit{1st}}$ $\textbf{\textit{Divergence}}$ $\textbf{\textit{Alert}}$ 
		$\tab$$\tab$$DeviationCounter++$  
		$\tab$ Wait for $\textbf{\textit{T}}_{\textit{\textbf{short}}}$
		$\tab$ $PrevS = S;$ $PrevB =  B$
		$\tab$ $UPDATE();$
		$\tab$ $\textbf{if}$ $(PrevS == S$ $\&\&$  $PrevB == B)$
		$\tab$$\tab$ $Send$ $\textbf{\textit{2st}}$ $\textbf{\textit{Divergence}}$ $\textbf{\textit{Alert}}$
		$\tab$$\tab$$DeviationCounter++$  
		$\tab$ $\textbf{else}$ 
		$\tab$$\tab$Go to line 9;
		$\tab$ $\textbf{end if;}$  
		$\tab$ Wait for $\textbf{\textit{T}}_{\textit{\textbf{long}}}$
		$\tab$ $PrevS = S;$ $PrevB =  B$
		$\tab$ $UPDATE();$
		$\tab$ $\textbf{if}$ $(PrevS == S$ $\&\&$  $PrevB == B)$
		$\tab$$\tab$ $Send$ $\textbf{\textit{Final}}$ $\textbf{\textit{Divergence}}$$\textbf{\textit{Alert}}$ 
		$\tab$$\tab$$DeviationCounter++$   
		$\tab$ $\textbf{else}$ 
		$\tab$$\tab$Go to line 9;
		$\tab$ $\textbf{end if;}$  
		$\textbf{else}$ 
		$\tab$$DeviationCounter=0$  
		$\tab$ Display $G \prime$ $\subseteq$ $G$ for the current patient state
		$\tab$ $\textbf{\textit{S}}= (S_{1},S_{2},\dots,S_{n})$
		$\textbf{end}$ $\textbf{if}$
		
		$UPDATE()$ $\{$
		$\tab$ $\textbf{for}$ $i$ = $1$ $to$ $n$
		$\tab$ $\tab$ $S_{i}= CurrentState(OA_{i})$
		$\tab$ $\tab$ $B_{i}= CurrentState(PA_{i})$
		$\tab$ $\textbf{end}$ $\textbf{for}$
		$\tab$ $S= (S_{1},S_{2},\dots,S_{n})$
		$\tab$ $B = (B_{1},B_{2},\dots,B_{n})$ $\}$ 
		\end{lstlisting}
		\caption{Convergence-Divergence Protocol}
		\label{code:select}
	}
	
\end{program}

\subsubsection{Best Practice Model}\label{sec:practice}

Best practice (manager) model codifies the well-established guidelines, categorized according to the patient state. We define the patient state $S$ as a tuple $(S_{1},S_{2},\dots,S_{n})$, in which $S_{i}$  is the current state in $ ith $ pathophysiological process and $n$ is the total number of these processes developed for the disease under study. Fo example, for cardiac arrest case we have developed \textit{arrhythmia}, \textit{blood gas imbalance (BGI)}, and \textit{renal insufficiency} automata representing three different pathophysiological processes involved in this disease. Therefore, $n=3$ and $S$ = $(VTach$, $Metabolic$ $Acidosis$, $Normal$ $Renal$ $Function)$ can be an example of the current state of a cardiac arrest patient. Best practice manager model periodically communicates with all the organ models to get their current states and determine the current patient state. As mentioned before, the encoded guidelines are categorized to patient state defined as explained above. Therefore, when the patient states changes, the best practice model determines the subset of guidelines for the current patient state. For cardiac arrest the ACLS algorithm by American Heart Association \cite{acls} is codified.

In addition to encoding the guidelines, we have included a \textit{divergence-convergence} protocol in the best practice model. As presented in the Section \ref{sec:physician}, the compositional model allows us to track and record the physician belief on the current patient state. However, to reduce the \textit{unintended} divergence from best practice, we propose the \textit{divergence-convergence} protocol. When there is a discrepancy between the organ automata and its corresponding physician automata an initial alert is generated to reduce any unintended deviation. If patient organ state continues to deteriorate or remains the same, two additional alerts are generated by the system. Note that all the timer-related values are configurable. After discussion with our physician collaborators, we designed the last alert to be triggered by a longer timer. When we transition to a new organ state and the best practice and physician diagnoses are still inconsistent, the timer routine restarts. However, if the patient state improves the system (by design) converges with the physician's decision. The protocol steps are presented in Pseudocode \ref{code:select}.  

Any deviation from best practice are logged by our system and may be used for additional review (can be specifically useful for residents). According to \cite{commonErrors}, \textit{failure to debrief after code} is one of common errors made in cardiac arrest care and the logging feature can help alleviate this situation. Furthermore, the logs can also be analyzed by the developers to improve the precision of the proposed models.

\subsection{Clinical Validation} \label{validation}

The executable organ models allow us to simulate the disease under study. The clinical simulation provides an environment for dynamic validation of the models, in addition to statically cross referencing the clinical requirements and the models. As mentioned before, we use an open source tool, called Yakindu, to develop the models. The tool provides an integrated modeling environment and allows simulation of the developed model to validate its dynamic behavior \cite{yakindu}. Physician can directly look at the organ models and cross reference with the specified clinical requirements. Note that, the developed models are at same level of abstraction the medical knowledge and reasoning are known to physician. If directly implemented in procedural style using a high level language such as Java, organ-centric patient state representation could not be easily validated by physicians. On the other hand, at our proposed level of abstraction, physicians with minimal IT background can easily understand the models and review the design. The following are a subset of clinical requirements validated for the developed organ models:

\begin{itemize}
	\item $CR_{1}$: All the pathophysiological processes relevant to the disease under study are modeled.
	\item $CR_{2}$: All the necessary insufficiency states are modeled for each pathophysiological process.
	\item $CR_{3}$: The correct and only necessary subset of physiological measurements are used to define the states in each automaton.
	\item $CR_{4}$: The definition of insufficiency states are correct.
	\item $CR_{5}$: Any changes in the physiological measurements will result in transition(s) to the correct insufficiency state(s).  
\end{itemize}

We have shown our developed models and simulation of cardiac arrest scenario to our physician collaborators for clinical validation. More details on our simulation of cardiac arrest is presented in Section \ref{sec:case-studies}. Note that for each pathophysiological process, the physician automaton has the exact same states as the best practice version of the corresponding automata. The transitions are simply enabled by the physician input. The correctness of the physician models are formally verified using UPPAAL (Section \ref{verification}).


\begin{figure*} [!t]
	\center
	\includegraphics[width=0.99\textwidth]{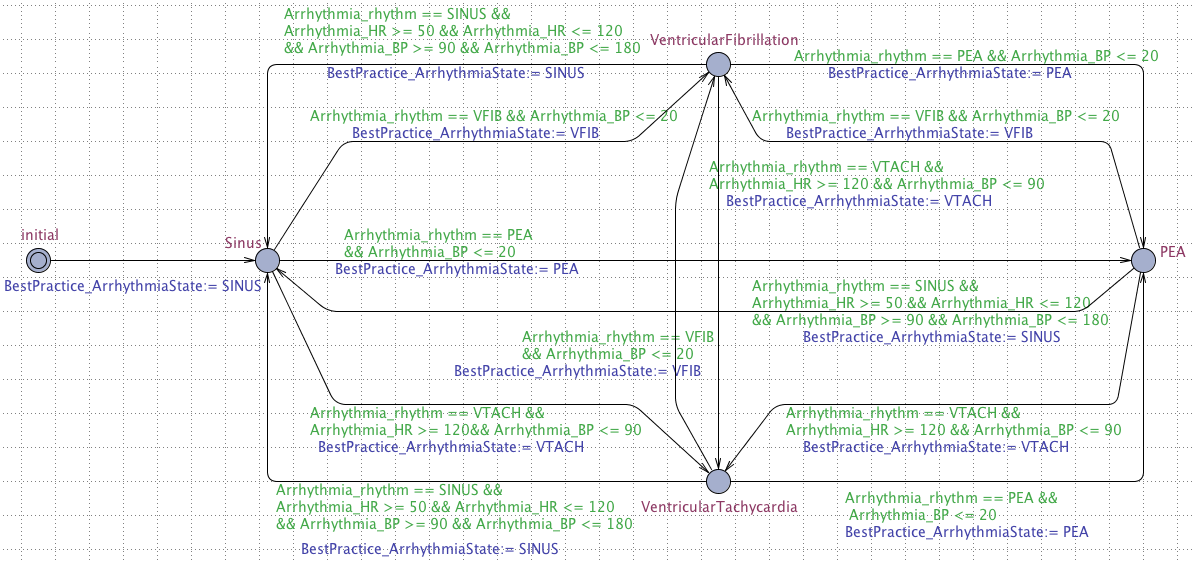} 
	\caption{Manually Translated Arrhythmia UPPAAL Model }
	\label{fig:ArrhythmiaUppaal}
\end{figure*}

\subsection{Formal Verification} \label{verification}

To formally verify the correctness of the models, we use UPPAAL model checker tool. Currently, the translation from Yakindu to UPPAAL is done manually. The correctness of the translation is a challenge that also needs to be addressed but is outside the scope of this paper. However, the UPPAAL model-checker is also based on the theory of timed automata and this make the translation process less complex. In \cite{Guo2016ICCPS} automatic translation of Yakindu models to UPPAAL is discussed. However, in this paper the translation is done manually until such research works mature.

As mentioned before, we use an open source tool, called Yakindu, to develop the models. The tool provides an integrated modeling environment and development based on the concept of statecharts. To make the traslation to UPPAAL less complex, we avoided using composite states, supported by statecharts, in the Yakindu models. UPPAAL model checker tool does not support composite states. Therefore, using them in our organ models would make the task of translation to UPPAAL more complex and possibly error-prone. Another reason for avoiding composite states is to maintain the model simplicity for clinical validation by physicians. Understanding complex computer science concepts, such as composite states, would be a non-trivial tasks for the physicians. Our goal is to keeps the models as simple and intuitive as possible. Note that we use the orthogonal regions supported in statecharts to represent a set of concurrent organ models, which we develop to track concurrent changes in the patient state. In UPPAAL, a system is also modeled as a network of communicating state machines. This makes the task of translation to UPPAAL and formal verification more straightforward.     

The Yakindu models may include state and transition actions. Since UPPAAL only supports transition actions, we also translated any necessary state action to a transition action. In Yakindu, the state actions include entry, exit and timer based actions. The entry action should be performed once when entering the state. We translated this to an action assigned to all the incoming transitions to the corresponding state. The exit action, which must be executed when existing the state, was translated to actions on all the outgoing transitions. For the timer-based state actions, we created a self-transition and used the timer as transition guard.

Figure \ref{fig:ArrhythmiaUppaal} shows the \textit{arrhythmia} UPPAAL model, as an example. The four states are mapped to four locations, each representing an arrhythmia state, with the exception of initial state.The entry state actions are translated to updates on all the incoming transitions to the corresponding states. For the Sinus state, an additional transition is added from the initial state to initialize the state variable.

A subset of presented clinical requirements are formalized and verified using UPPAAL. Some clinical requirements such as $CR_{1}$ may only validated by a physician. We have translated our cardiac arrest Yakindu models to UPPAAL and formalized these clinical requirements and several system requirements as (timed) computation tree logic, CTL, formulas. The exact formula depends on the disease under study. A subset of verified properties are shown in Table \ref{table:properties1}. The absence of deadlock in the model is formalized as $P1$ and $P2$ is an example of reachability properties ($P2$). The safety properties $P3$ and $P4$ are examples of formalized $CR_{4}$ and $CR_{5}$ for the arrhythmia model.

\begin{table}
	{\scriptsize
		\begin{tabular} {| c | p{6.6cm}|}
			\hline
			\bf{Property} & \bf{Fromula}\\ \hline
			P1 & $A[]$ $Not$ $Deadlock$\\ \hline
			P2 & $E<>$ $ArrhythmiaBP.VentricularFibrillation$  \\ \hline
			P3 & $A<>$ $((Rhythm == VFIB$ \&\& $BP \leq 20)$ $\implies $ $ArrhythmiaBP.VentricularFibrillation)$\\ \hline
			P4 & $A<>$ $(ArrhythmiaBP.VentricularFibrillation$ $\implies$ $(Rhythm == VFIB$ \&\& $BP\leq 20))$ \\ \hline
		\end{tabular}
	}
	\caption{Property List}
	\label{table:properties1}
\end{table}

\subsection{Guidance System Integration} \label{sec:guidance}

As shown in Figure \ref{fig:approach}, we can integrate the developed models with an existing guidance system The Java code can be generated automatically from the Yakindu model and integrated with the existing system using some glue code. However, the tool also allows direct interaction between the Yakindu model and existing Java code. Therefore, for visualization purposes we integrated the cardiac arrest model with an existing guidance system, previously developed by our research group. The models interact with the existing system using our handwritten communication code. We specifically interact with two components of this existing system: a \textit{patient control panel} and \textit{graphical display}, which allow us to simulate different clinical scenarios. 

The communication follows the interface shown in Figure \ref{fig:overview}. The patient measurements is provided to the model using \textit{patient control panel}, which will trigger each pathophysiological state machine to transition to the states corresponding to the current measurements. The \textit{"suspected"} organ states are then sent to the \textit{main display} for the attending physician to see. 
We have emulated the physician response using a set of Yakindu events.  Physician can enter its response to the new suspected patient states. The responses will be received and recorded by our physician model. If the new patient state is not confirmed or a different organ state is entered by the physician, the convergence-divergence protocol, as explained in Section \ref{sec:practice}, is executed. However, if the new patient state is confirmed, the new guidelines are sent to the \textit{main display}. More details on the clinical simulation of cardiac arrest guidance system is presented in Section \ref{sec:case-studies}.

\section{Cardiac Arrest Case Study} \label{sec:case-studies}

A brief discussion of the cardiac arrest scenario is presented in Section \ref{sec:model}. We discussed the set of organ insufficiency models for cardiac arrest and elaborated the \textit{arrhythmia} model as an example (Figure \ref{fig:arrhythmiaBP}). The cardiac arrest model includes two additional pathophysiological processes: \textit{blood gas imbalance (BGI)}, and \textit{renal insufficiency}. Due to space limitation, the details on developed automata is presented in the appendix.

The best practice model for the cardiac arrest scenario codifies the the ACLS algorithm by American Heart Association \cite{acls}, and other well-established guidelines recommended by our physician collaborators. Readers are referred to the appendix for more details on ACLS algorithm. In addition to the these guidelines, the cardiac arrest best practice model includes the divergence-convergence protocol, as explained in Section \ref{sec:practice}.

\begin{figure*}[!t] 
	\centering
	\subfigure[Simulation Snapshot]{\label{fig:warning} \includegraphics[width=0.7\textwidth]{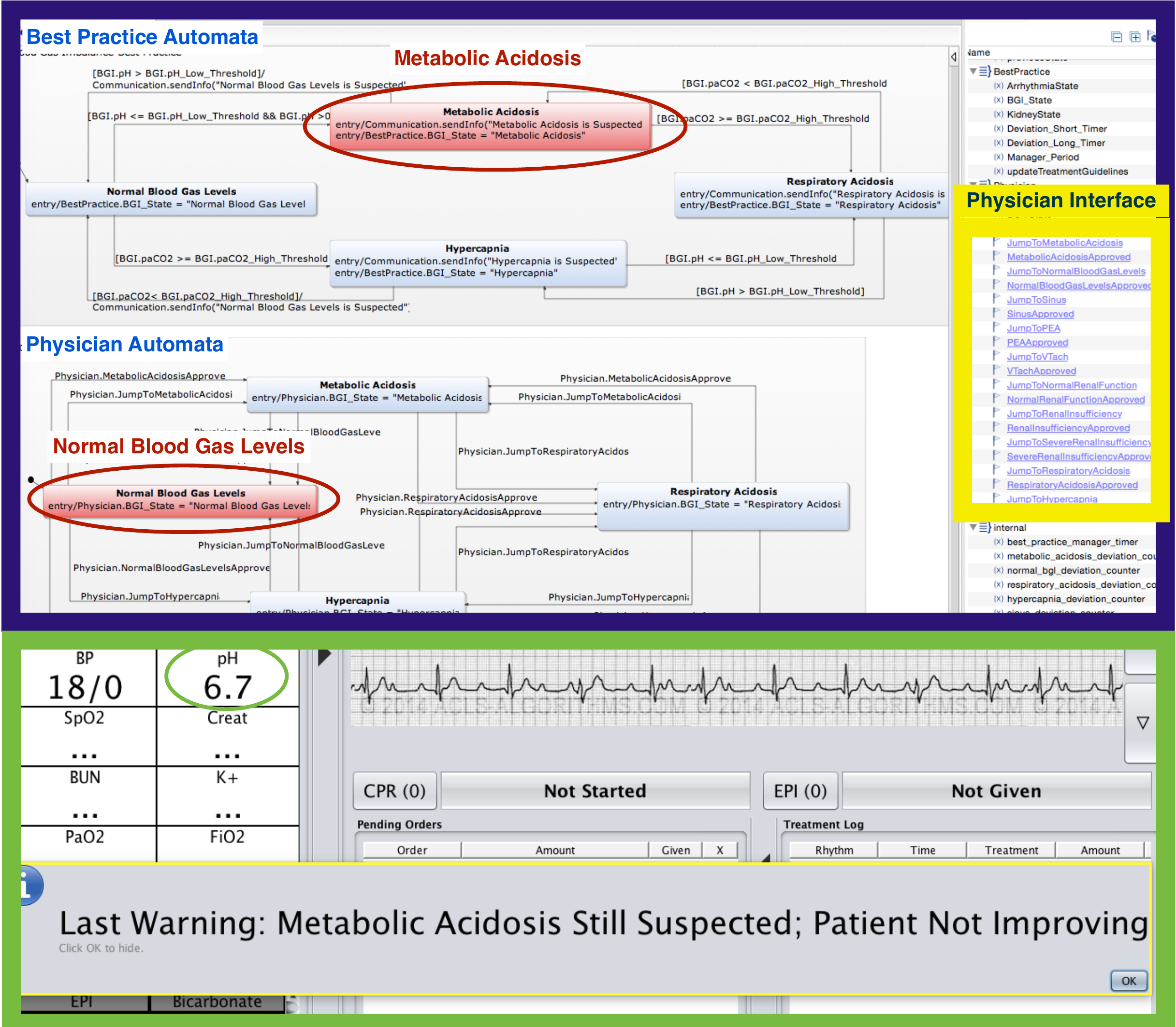}}
	\subfigure[Log Snapshot] {\label{fig:log} \includegraphics[width=0.8\textwidth]{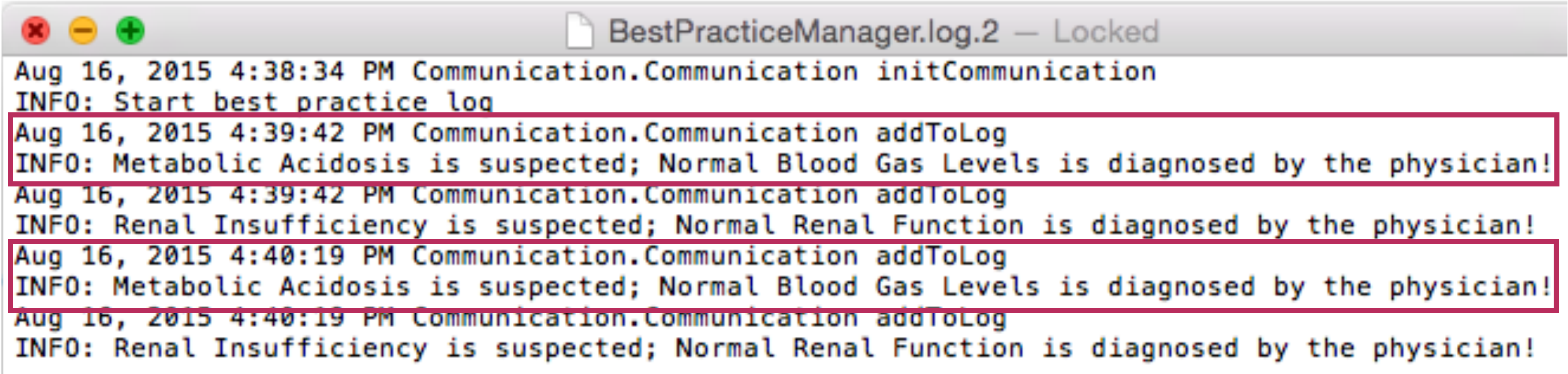}}
	\caption{Inconsistent Physician and Best Practice Blood Gas Imbalance Automata}
	\label{fig:inconsistent}
\end{figure*}  

\begin{figure} 
	\center
	\includegraphics[width=0.9\columnwidth]{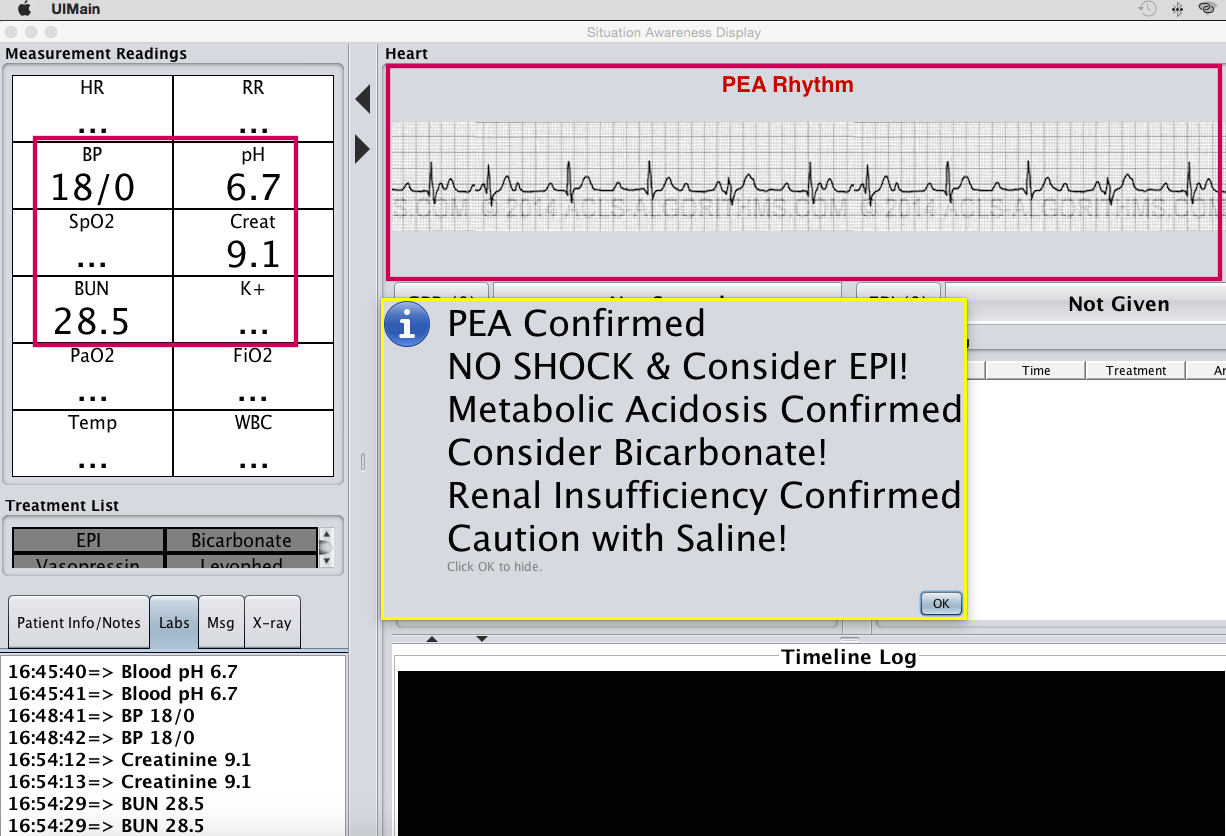} 
	\caption{Best Practice Guidelines Generated by our System}
	\label{fig:consistent}
\end{figure}

To further demonstrate how the models can be used to design clinical systems, we have developed a cardiac arrest guidance system using our constructed models and following the approach presented in Section \ref{sec:guidance}. Figure \ref{fig:warning} shows a simulation snapshot including the organ models (Statechart implementation) and the graphical display. The snapshot shows a point in the simulation, in which the patient's blood $pH$ value indicates metabolic acidosis. However, the physician does not approve the organ state suggested by the system. Therefore, the two versions of Blood Gas Imbalance automata have different runtime states. We have defined a set of events in Yakindu to emulate the physician's response to the system's suggestions. As shown in Figure \ref{fig:log}, the deviation from best practice guided acidosis diagnosis is logged by the system. Note that when blood $pH$ goes above the threshold, the best practice \textit{BGI} automaton automatically transitions to \textit{Normal Blood Gas Levels} state and converges with the physician diagnosis. This case is not shown due to space limitations.   

When the physician approves the organ states suggested by the system, the subset of best practice guidelines applicable to the current patient state are placed on the main display. Figure \ref{fig:consistent} shows a simulation snapshot, in which $PEA$ rhythm, metabolic acidosis and renal insufficiency states have been approved by physician. We have shown our simulation of cardiac arrest guidance system to our physician collaborators to ensure the clinical correctness. The following are a subset of clinical requirements for the cardiac arrest, which are instantiations of the generic requirements presented in Section \ref{validation}.  

\begin{itemize}
	\item $CR_{1}$: All the pathophysiological processes relevant to cardiac arrest are modeled.
	\item $CR_{2}$: All the abnormal rhythms relevant to cardiac arrest are modeled in the Arrhythmia machine.
	\item $CR_{3}$: EKG rhythm, heart rate and blood pressure are the correct and necessary physiological measurements to define different arrhythmia states.
	\item $CR_{4}$: The definition of each arrhythmia, BGI and renal state is correct.
	\item $CR_{5}$: Any changes in blood $pH$ and/or $PaCO2$ result in transition to the correct blood gas imbalance state.    
\end{itemize}


After validating the clinical requirements, using Yakindu simulation, we verify a set of formalized safety properties using UPPAAL model checker. We followed the approach explained in Section \ref{verification}, and manually translated all the models to UPPAAL. Due to space limitation, the other UPPAAL models are not presented in this paper. Some of the properties have been presented in Section \ref{verification}. Additional verified properties are listed in Table \ref{table:properties2}. For example, $P7$ ensures the correctness of physician model and $P8$ is a safety property of the divergence-convergence protocol.

\begin{table}
	{\scriptsize
		\begin{tabular} {| c | p{7.0cm}|}
			\hline
			\bf{Property} & \bf{Fromula}\\ \hline
			
			P5 & E<> $BGI\_BestPractice.MetabolicAcidosis$  \\ \hline
			P6 & A<> $BGI\_BestPractice.MetabolicAcidosis$ $\implies$ $(pH < PH\_LOW\_THRESHOLD)$ \\ \hline
			P7 & A[] $BGI\_Physician.MetabolicAcidosis \implies Physician\_BGI\_State == METABOLIC\_ACIDOSIS$ \\ \hline
			P8 & A<> $(BGI\_BestPractice.MetabolicAcidosis$ \&\&
			$BGI\_Physician.NormalBloodGasLevels)$ $\implies $ $(MetabolicAcidosis\_DeviationCounter >0)$\\ \hline
		\end{tabular}
	}
	\caption{Cardiac Arrest Property List}
	\label{table:properties2}
\end{table}

\section{Related Work}\label{sec:related} 

The necessity of proper infrastructures and the principles of medical systems have been discussed in \cite{high-confidence, hatcliff2012rationale}. Medical Device Coordination Framework (MDCF) facilitates integration of different medical devices, and provides an app programming environment for easy assembling of a new medical application \cite{open-testbed, MDCF}. Our focus is on developing executable disease models to use for designing an integrated monitoring system for acute care. The proposed system can benefit from the services provided by such platforms.


Most current medical monitoring devices and systems provide single (physiological) measurement-alerts and thus little contextual information to medical staff. Some research has been done on exploiting the relations between measurements and the development of smart alarms \cite{imhoff2006alarm}, \cite{imhoff2009smart}. However, we believe that an integrated computerized system is necessary to help medical staff keep track of critical changes in patient state and perform treatments in a timely manner. 

A great body of work in this domain has been focused on expert systems, which are designed to provide medical diagnosis and treatment suggestions by codifying experts' knowledge, reasoning using the rules and deducing new knowledge \cite{hudson2006medical}. Several inference and decision making methodologies have been developed, such as rule-based reasoning \cite{tsumoto1996automated}, fuzzy logic \cite{hudson1994fuzzy} and probabilistic network \cite{andreassen1991medical}. Approaches based on machine learning aim to excerpt common patterns from empirical medical data and make decisions based on the learned behaviors \cite{kononenko2001machine}. However, most clinical problems are complicated and involve many dynamic decision-making steps; therefore, straightforward attempts to chain together larger sets of rules encounter major difficulties \cite{szolovits1988artificial}. Unlike such decision support systems, in this paper, we focus on developing an organ-centric patient monitoring system to aid medical staff to keep track of critical changes in the patient state and adhere to best practice guidelines. 

Our computational approach codifies patient information into executable organ automata.The main reason behind developing organ-centric disease models is that medical literature and training represent disease as a set of abnormal organ states. There have been similar works on organization of patient data. The authors in \cite{sharp} propose a framework to transform EHR data into standards-conforming, comparable information suitable for large-scale analysis and inference. In \cite{bodySystem} a new application system is proposed to store and categorize patient information according to body system view. However, the focus is on storing and displaying the value and trend of correlated (raw) measurements together. We focus on developing executable disease models that can be directly used to derive the best practice assist system.

\section{Concluding Remarks}\label{sec:conclusion} 

In this paper, we propose a domain-specific model-driven design approach for designing clinical guidance systems. We introduce an organ-centric paradigm to construct clinical models, and also develop a physician model to track physician-system interactions. We translate the model into timed automata, based on which, we formalize a set of clinical and system safety requirements as computation tree logic formulas and use the UPPAAL model checking tool to formally verify them. As part of our future work, we will consider more complex diseases such as sepsis and discuss the use of physiological trends in organ models. Furthermore, to demonstrate the effectiveness of our system, we are in the process of evaluating our models and integrated system in the context of simulated resuscitation using an ACLS training scenario.

\section*{Acknowledgment}
This work is supported by NSF CNS 1329886, NSF CNS 1545002, NIH Sub MGH 217183 and in part by ONR N00014-14-1-0717.

\bibliographystyle{abbrv}
\bibliography{ICCPS}


\begin{figure*}[!t] 
	\centering
	\subfigure[Best Practice Automaton]{\label{fig:bgiBP} 
	\includegraphics[width=0.99\textwidth]{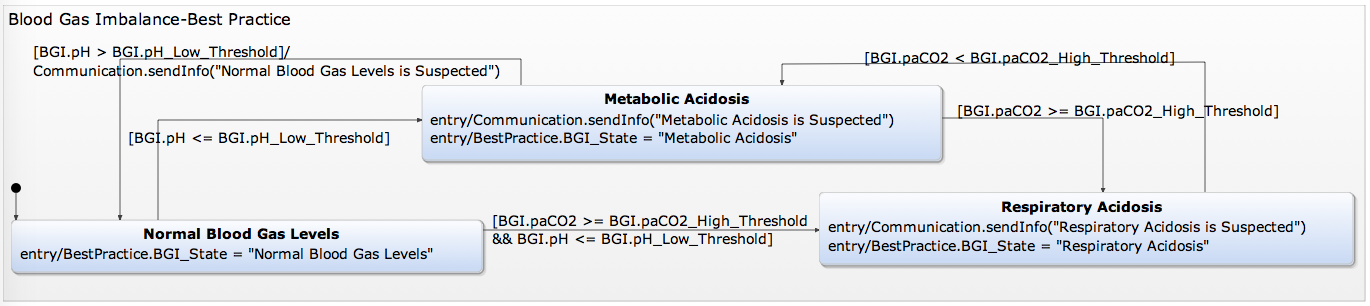}}
	\subfigure[Physician Automaton] {\label{fig:bgiPhysician} 		
	\includegraphics[width=0.99\textwidth]{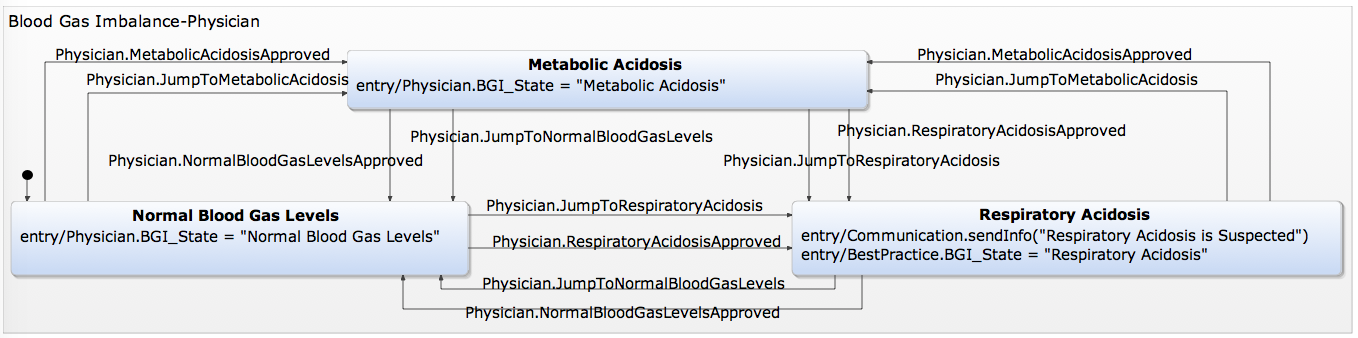}}
	\caption{Blood Gas Imbalance Automata}
	\label{fig:BGI}
\end{figure*}  


\begin{figure*}[!t] 
	\centering
	\subfigure[Best Practice Automaton]{\label{fig:kidneyBP} 	
	\includegraphics[width=0.99\textwidth]{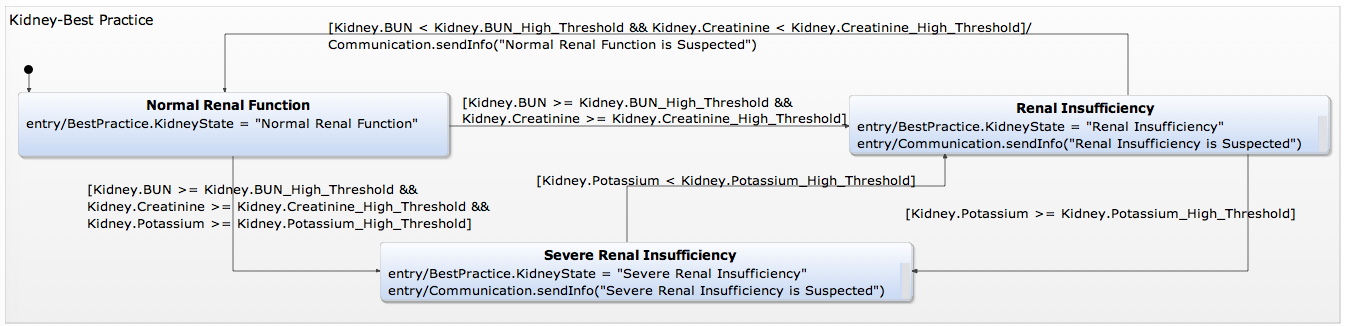}}
	\subfigure[Physician Automaton] {\label{fig:kidneyPhysician} 		
	\includegraphics[width=0.99\textwidth]{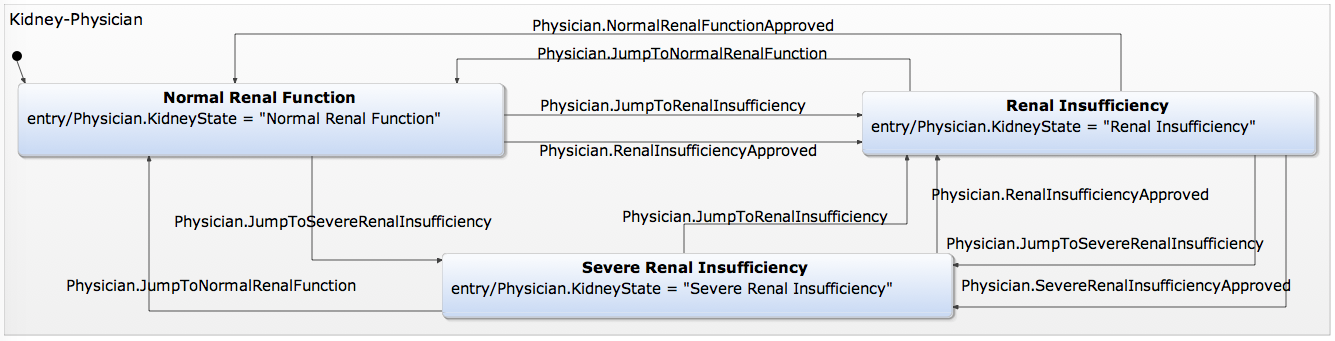}}
	\caption{Kidney Automata}
	\label{fig:kidney}
\end{figure*}  

\appendix
\section{Cardiac Arrest Yakindu Models}

In this section we present the \textit{blood gas imbalance (BGI)}, and \textit{renal insufficiency} automata (Figures \ref{fig:bgiBP}, and \ref{fig:kidneyBP}). The nodes represent different stages/types of imbalance in arterial blood gas levels and renal insufficiency according to existing medical knowledge. The changes in the relevant physiological measurements and lab values, which result in satisfaction of the condition for a new organ state cause the transition. For example, when the blood $pH$ value from a recent ABG test is below the lower threshold, the \textit{blood gas imbalance} automaton transitions to \textit{metabolic acidosis}. Similarly, the renal insufficiency automaton, shown in Figure ref{fig:kidneyBP}), keeps track of current renal function state, according to best practice monitoring guidelines. 

\end{document}